%% file: mis.tex
\newcommand{\short}[1]{#1}     
\newcommand{\extended}[1]{}    

\documentclass[submission,copyright,creativecommons]{eptcs}
\providecommand{\event}{Gandalf 2013} 

\usepackage{tabularx}
\usepackage{latexsym}
\usepackage{amssymb}
\usepackage{color}
\usepackage[utf8x]{inputenc}
\usepackage{amsmath}
\usepackage{graphics,graphicx}
\usepackage{wojtek11}
\usepackage{theoremsdecl}
\usepackage[mathscr]{eucal}

\def\Act{\mathit{Act}}
\def\Agtnames{\mathit{Agtnames}}
\def\In{{\mathcal{I}\!\!\mathit{n}}}
\def\Tok{\mathcal{T}\!\!\!\mathit{ok}}

\def\NN{{\mathbb{N}}}

\def\pSet{\mathcal{P}}

\newcommand{\agent}{\pmb{a}\xspace}
\newcommand{\allsets}[1]{\mathcal{P}({#1})}
\newcommand{\allbags}[1]{\mathcal{B}({#1})}
\newcommand{\complFn}{\mathbb{N}\rightarrow\mathbb{R}_+\cup\{0\}}

\newcommand{\KolorAM}[1]{#1}
\newcommand{\noteWJ}[1]{\textcolor{darkgreen}{#1}}

\newcommand{\Meski}{M{\c{e}}ski\xspace}

\title{Modularity and Openness in Modeling Multi-Agent Systems}
\author{Wojciech Jamroga
\institute{Computer Science and Communication, and \\ Interdisciplinary Centre on Security, Reliability and Trust \\ University of Luxembourg}
\email{wojtek.jamroga@uni.lu}
\and
Artur \Meski
\institute{Institute of Computer Science \\ Polish
Academy of Sciences, Warsaw, Poland, \\
and FMCS, University of \L{}\'{o}d\'{z}, Poland}
\email{meski@ipipan.waw.pl}
\and
Maciej Szreter
\institute{Institute of Computer Science \\ Polish
Academy of Sciences, Warsaw, Poland}
\email{mszreter@ipipan.waw.pl}
}
\def\titlerunning{Modularity and Openness in Modeling Multi-Agent Systems}
\def\authorrunning{W. Jamroga, A. \Meski \& M. Szreter}

\begin{document}
\maketitle

\begin{abstract}
We revisit the formalism of modular interpreted systems (MIS) which encourages modular and open modeling of synchronous multi-agent systems.  The original formulation of MIS did not live entirely up to its promise.
In this paper, we propose how to improve modularity and openness of MIS by changing the structure of interference functions.
These relatively small changes allow for surprisingly high flexibility when modeling actual multi-agent systems. We demonstrate this on two well-known examples, namely the trains, tunnel and controller, and the dining cryptographers.

Perhaps more importantly, we propose how the notions of multi-agency and openness, crucial for multi-agent systems, can be precisely defined based on their MIS representations.
\end{abstract}

\input{10_intro}

\input{21_mis_defs}

\input{22_mis_exmpl}

\input{30_agentivity}
\input{40_openness}

\input{50_conclusions}

\bibliographystyle{eptcs}
\bibliography{wojtek-own,wojtek}

\end{document}

%% file: 10_intro.tex
\section{Introduction}

The paradigm of multi-agent systems (MAS)\extended{~\cite{Weiss99mas,Wooldridge02intromas}} focuses on systems consisting of
autonomous entities acting in a common environment. Regardless of whether we
deem the entities to be intelligent or not, proactive or reactive, etc., there
are two design-level properties that a multi-agent system should satisfy.
First, it should be \emph{modular} in the sense that it is inhabited by
\emph{loosely coupled} components. That is, interaction between agents is
crucial for the system, but it should be relatively scarce compared to the
intensity of local computation within agents (otherwise the system is in fact
a single-agent system in disguise). Secondly, it should be \emph{open} in the
sense that an agent should be able join or leave the system without changing
the design of the other components.

Models and representations of MAS can be roughly divided into two
classes.\extended{
  \footnote{
    In this paper, we are interested in modeling MAS in the
    context of formal analysis and reasoning. Thus, we focus on models and
    representations that support verification and reasoning in a suitable logic.
    That leaves e.g.\ modeling paradigms of agent-oriented software engineering
    out of the scope of the paper. }
} 
On one hand, there are models of various agent logics, most notably modal logics of
knowledge\extended{~\cite{Kripke63semantical}}, action\extended{~\cite{Harel00dynamic}}, time\extended{~\cite{Pnueli77temporal,Emerson82sometimes}}, and strategic ability~\cite{Emerson90temporal,Fagin95knowledge,Alur02ATL}. These models are well suited for theoretical
analysis of general properties of agent systems. However, they are too abstract in the
sense that: (a) they are based on abstract notions of global state and global
transition so the structure of a model does not reflect the structure of a MAS
at all, and (b) they come with neither explicit nor implicit methodology for
design and analysis of actual agent systems.
At the other extreme there are practical-purpose high-level representation
languages like Promela~\cite{Holzmannn97spin},
Estelle~\cite{Dembinski03verics}, and Reactive Modules (RM)~\cite{DBLP:journals/fmsd/AlurH99b}.
They are application-oriented, and usually include too many features to be
convenient for theoretical analysis.  The middle ground consists of formalisms
that originate from abstract logical models but try to encapsulate a
particular modeling methodology. For instance, interpreted
systems~\cite{Fagin95knowledge} support local design of the state space;
however, transitions are still global, i.e., they are defined between global
rather than local states.  Synchronous automata
networks~\cite{Gecseg86automata} and ISPL
specifications~\cite{Lomuscio06mcmas,Raimondi06phd} push the idea further:
they are based on local states and semi-local transitions, i.e., the outcome
of a transition is local, but its domain global\extended{ (transitions in ISPL are
functions from global states to local states)}.  This makes agents hard to separate from one another in a
model, which hampers its modularity.  On the other hand, concurrent
programs~\cite{Lichtenstein85checking} and asynchronous automata
networks~\cite{Gecseg86automata} are fairly modular but they support only
systems whose execution can be
appropriately modeled by interleaving of local actions and/or events.

Modular Interpreted Systems (MIS) are a class of models proposed
in~\cite{Jamroga07mis-aamas} to achieve separation of the
interference between agents from the local processing within agents. The main
idea behind MIS was to encapsulate the way agents' actions interfere by so
called \emph{interaction tokens} from a given alphabet $\In$, together with
functions $out_i,in_i$ that define the \emph{interface} of agent $i$. That is,
$out_i$ specifies how $i$'s actions influence the evolution of the other
agents, whereas $in_i$ specifies how what rest of the world influences
the local transition of $i$. Modular interpreted systems received relatively
little attention, though some work was done on studying computational
properties of the related verification problem~\cite{Jamroga06mis-tr},
facilitating verification by abstraction~\cite{Koester11abstraction}, and
using MIS to analyze homogeneous multi-agent systems~\cite{Calta12phd}.  This
possibly stems from the fact that, in their original incarnation, MIS are not
as modular and open as one would expect. More precisely, the types of
functions used to define interference fix the number of agents in the MIS.
Moreover, the assumption that all the functions used in a model are
deterministic limit the practical applicability, as modeling of many natural
scenarios becomes cumbersome.

In this paper, we try to revive MIS as an interesting formalism for modeling
multi-agent systems.  We propose how to improve modularity and openness of the
original class by changing the structure of interference functions $out,in$.
The idea is to use multisets of interference tokens instead of $k$-tuples.
This way, we do not need to ``hardwire'' information about other modules
inside a module.
\extended{
  Of course, there are infinitely many multisets that draw
  elements from a set even if the set is finite. Thus, our interference
  functions become infinite and need finite representations. We represent them
  with decision lists in a similar way to \emph{guarded transitions} in implicit
  concurrent game systems~\cite{Laroussinie08expATL} and ISPL
  specifications~\cite{Raimondi06phd}.
}
Additionally, we assume that the ``manifestation'' function $out$ can be nondeterministic.  These relatively
small changes allow for surprisingly high flexibility when modeling MAS. We
demonstrate that on two well-known benchmark examples: trains, tunnel and
controller, and the dining cryptographers.

Perhaps more importantly, we propose how two important features of multi-agent systems can be formally defined,
based on MIS representations. First, we show how to decide if a system
is designed in a proper multi-agent way by looking at the relation between the
complexity of its interference layer to the complexity of its
global unfolding.
%
Moreover, we define the degree of openness of a MIS as the complexity of the minimal
transformation that the model must undergo in order to add a new agent to the
system, or remove an existing one. We apply the definitions to our benchmark models, and show that different variants of cryptographers grossly differ in the amount of openness that they offer.

The paper has the following structure. In Section \ref{sec:mis}, the new variant of
MIS is defined, along with its execution semantics. Section
\ref{sec:modeling} presents MIS representations for two benchmarks: Tunnel, Trains and Controller (TTC)
and Dining Cryptographers (DC). A graphical notation is provided to make the
examples easier to read. In Sections \ref{sec:modularity} and
\ref{sec:openness}, we propose formal definitions of multi-agency and openness,
respectively, and apply them to several variants of the benchmarks. Section
\ref{sec:conclusions} concludes the paper.


\subsection{Related Work}

The modeling structures discussed in this paper share many similarities with existing modeling frameworks, in particular with Reactive Modules~\cite{DBLP:journals/fmsd/AlurH99b}. Still, MIS and RMs have different perspectives: Reactive Modules is an application-oriented language, while the focus of modular interpreted systems is more theoretic. This results in a higher abstraction level of MIS which are based on abstract states and interaction tokens. MIS aim at separating internal activities of modules and interactions between modules, what is not (explicitly) featured in RM.

Modularity in models and model checking has been the focus of many papers. Most notably, Hierarchical State Machines of Alur et al.~\cite{Alur99communicatingHSM,Torre08HSM} and the approach of hierarchical module checking by Murano et al.~\cite{Murano08hierarchical} feature both ``horizontal'' and ``vertical'' modularity, i.e., a system can be constructed by means of parallel composition as well as nesting of modules.
Similarly, dynamic modifications and ``true openness'' of models has been advocated in~\cite{DBLP:conf/concur/FisherHNPSV11}. In that paper, Dynamic Reactive Modules (DRM) were proposed, which allow for dynamic reconfiguration of the structure of the system (including adding and removing modules).
Our approach differs from the ones cited above in two ways. On one hand, we focus on an abstract formulation of the \emph{separation of concerns} between modules (and agents), rather than providing concrete mechanisms that implement the separation. On the other, we define indicators that show \emph{how good the resulting models is}. That is, our measures of agentivity and openness are meant to assess the model ``from the outside''.
In particular, the focus of the DRM is on providing a mechanism for adding and removing agents in the RM representation. We implement these operations on the meta-level, as a basis of the mathematical measure of openness. Our work could in principle be applied to DRMs and other formalisms, but it would require defining the appropriate multi-agent mechanisms which are already present in Interpreted Systems.

%% file: 21_mis_defs.tex
\def\Controller{{ctrl}}

\section{Modular Interpreted Systems Revisited}\label{sec:mis}

Modular interpreted systems were proposed in~\cite{Jamroga07mis-aamas} to encourage {modular} and {open} design of {synchronous} agent systems. Below, we present an update on the formalism.
The new version of MIS differs from the original one~\cite{Jamroga07mis-aamas} as follows. First, a single agent can be now modeled by more than one module to allow for compact design of agents' local state spaces and transition functions. Secondly, the type of function $in_i$ is now independent from the structure and cardinality of the set of agents, thus removing the main obstacle to modularity and openness of representation in the previous version.
Thirdly, the interaction functions $in_i,out_i$ are nondeterministic in order to enable nondeterministic choice and randomization (needed, e.g., to obtain fair scheduling or secure exchange of information). Fourthly, we separate agents from their names. This way, agents that are not present in the ``current'' MIS can be referenced in order to facilitate possible future expansion of the MIS.

\subsection{New Definition of MIS}\label{sec:mis-def}

Let a bag (multiset) over set $X$ be any function $X \to \NN$. The set of all bags over $X$ will be denoted by
$\allbags{X}$, and the union of bags by $\uplus$.
\begin{definition}[Modular interpreted system]
\label{def:new-mis}
We define a modular interpreted system (MIS) as a tuple

\smallskip
\centerline{$S = (\Agtnames,\Act,\In,\Agt),$}

\smallskip
\noindent
where
$\Agtnames$ is a finite set of \emph{agent names},
$\Act$ is a finite set of \emph{action names},
$\In$ is a finite \emph{interaction alphabet},
and $\Agt = \{ a_1, \ldots, a_k \}$ is a finite set of \emph{agents}
(whose structure is defined in the following paragraph).
A set of {\em directed tokens}, used to specify the recipients of
interactions, is defined as $\Tok = \In \times (\Agtnames \cup \{\epsilon\})$,
where $\epsilon$ denotes that the interaction needs to
be broadcasted to all the agents in the system.

Each agent $a_j = (id,\set{m_1,\dots,m_n})$ consists of a unique name
$id\in\Agtnames$ \KolorAM{(also denoted with $name(a_j)$)}, and one or more \emph{modules}
$m_j = (St_j, \KolorAM{Init_j}, d_j, out_j, in_j, o_j, \Pi_j, \pi_j)$, where:
\begin{itemize2}
\item $St_j$ is a set of \emph{local states},
\item \KolorAM{$Init_j \subseteq St_j$ is the set of initial states},
\item $d_j : St_j \to \pSet(Act)$ defines local availability of actions; for convenience of the notation,
  we additionally define the set of \emph{situated actions} as \
  $D_j = \{(q_j, \alpha) \mid q_j \in St_j, \alpha \in d_j(q_j) \}$,
\item $out_j$, $in_j$ are \emph{interaction} (or \emph{interference}) \emph{functions}:
  \begin{itemize2}
  \item $out_j : D_j \to \allsets{\allsets{\Tok}}$ refers to the set of
    influences (chosen nondeterministically)
	that a given situated action (of module~$m_j$)
	may possibly have on the
	recipients of the embedded interaction symbols, and
  \item $in_j : St_j \times \allbags{\In} \to \pSet(\In)$ translates
	external manifestations from the other modules
    into the (nondeterministically chosen) ``impression'' that they make on
	module $m_j$ depending on the local state of $m_j$; we assume $in_j(\cdot) \neq
	\emptyset$;
  \end{itemize2}
\item $o_j : D_j \times \In \to \pSet(St_j)$ is a \emph{local transition function} (possibly nondeterministic),
\item $\Pi_j$ is a set of \emph{local propositions} of module $m_j$ (we require
  that $\Pi_j$ and $\Pi_m$ are disjoint when $j\neq m$),
\item $\pi_j: \Pi_j \to \pSet(St_j)$ is a \emph{valuation} of these propositions.
\end{itemize2}
Additionally, we define the \emph{cardinality of $S$} (denoted $card(S)$) as the number of agents in $S$.
\end{definition}

Typically, each agent in a MIS consists of exactly one module, and we will use
the terms interchangeably. Also, we will omit $Init_j$ from the description of a module whenever $Init_j = \States_j$.

Note that function $in_j$ is in general infinite. For practical purposes, finite representation of $in_j$ is needed. We use decision lists similarly to~\cite{Laroussinie08expATL,Raimondi06phd}.
Thus, $in_i$ will be described as an ordered list of pairs of the form $condition \mapsto value$.
The first pair on the list with a matching condition decides on the value of
the function. The conditions are boolean combinations of membership and cardinality tests, and are defined over the
variable $s$ for the conditions defined on states, and over $H$ for the
conditions on multisets of received interferences.
We require that the last condition on the list is
$\top$, so that the function is total.
Several examples of MIS's are presented in Sections~\ref{sec:modeling} and~\ref{sec:openness}.

\subsection{Execution Semantics for MIS}

\def\cAgt{\mathcal{A}}
\def\cSt{\mathcal{S}\!t}
\def\cPV{{PV}}
\def\cPVf{\mathcal{V}}
\def\cAct{{\mathcal{A}\!\!ct}}
\def\cProt{\mathfrak{d}}
\def\cTF{t}
\begin{definition}[Explicit models]
A \emph{nondeterministic concurrent epistemic game structure} \emph{(NCEGS)} is a tuple \
%
%
$C = (\cAgt, \cSt, \KolorAM{\cSt_0}, \cPV, \cPVf, \cAct, \cProt, \cTF, \sim_1, \dots, \sim_k)$,\
where:
\  $\cAgt = \{1, \ldots, k\}$ is a nonempty set of agents,
\  $\cSt$ is a nonempty set of states,
\  \KolorAM{$\cSt_0 \subseteq \cSt$ is the set of initial states,}
\  $\cPV$ is a set of atomic propositions,
\  $\cPVf : \cPV \to \allsets{\cSt}$ is a valuation function,
\  $\cProt : \cAgt \times \cSt \to \allsets{\cAct}$ assigns nonempty sets of actions
	available at each state, and
\  $\cTF$ is a (nondeterministic) transition function
	that assigns a nonempty set $Q = \cTF(q, \alpha_1, \ldots, \alpha_k)$ of outcome states
	to a state~$q$, and a tuple of actions $(\alpha_1, \ldots, \alpha_k)$
	that can be executed in $q$.
%
%
\end{definition}
We define the semantics of MIS through an unfolding to NCEGS.
\begin{definition}[Unfolding of MIS]
\label{def:unfolding}
%
%
Unfolding of the modular interpreted system S from Definition~\ref{def:new-mis} to a nondeterministic concurrent epistemic
game structure\ $NCEGS(S) = (\cAgt', \cSt', \KolorAM{\cSt_0'}, \cPV', \cPVf', \cAct', \cProt', \cTF')$\ is defined
as follows:
\begin{itemize2}
\item $\cAgt' = \{1, \ldots, k\}$, and $\cAct' = \Act$,
\item $\cSt' = \prod_{i=1}^{k} St_i$,
\item \KolorAM{$\cSt_0' = \{ (q_1,\dots,q_k) \mid (\forall i\in\{1,\dots,k\})~
q_i \in Init_i \}$,}
\item $\cPV' = \bigcup_{i=1}^{k} \Pi_i$,
	and $\cPVf'(p) = \pi_i(p)$ when $p \in \Pi_i$,
\item $\cProt'(i,q) = d_i(q_i)$ for global state $q = (q_1, \ldots, q_k)$, and
$i\in\cAgt'$,
\item The transition function $\cTF'$ is constructed as follows.
	Let $q = (q_1, \ldots, q_k)$ be a state,
	and $\alpha = (\alpha_1, \ldots, \alpha_k)$ be a joint action.
	We define an auxiliary function $\mathfrak{oi}_i(q_i,\alpha_i)$
	of all the possible interferences of agent $i$,
	for $q_i$, and $\alpha_i$:
	$\gamma' \in \mathfrak{oi}_i(q_i, \alpha_i)$
	iff
	there exist $T_1$, \ldots, $T_k$ such that
	$T_j \in out_j((q_j, \alpha_j))$, and
	$\gamma' \in in_i(q_i, \mathcal{I}_1 \uplus \ldots \uplus \mathcal{I}_k)$,
	\KolorAM{where
	$\mathcal{I}_j =
	\{ \gamma_j \mid (\exists r \in \{ name(a_j), \epsilon \})~(\gamma_j, r) \in T_j \}$
  for all $j \in \cAgt'$.}

	Then $(q'_1, \ldots, q'_k) \in \cTF(q, \alpha_1, \ldots, \alpha_k)$
	iff $q'_i \in o_i((q_i, \alpha_i), \gamma)$,
	where $\gamma \in \mathfrak{oi}_i(q_i, \alpha_i)$;
\item $q \sim_i q'$ iff $q$ and $q'$ agree on the local states of all the modules in agent $a_i$.
\end{itemize2}
\end{definition}

Definition~\ref{def:unfolding} immediately provides some important logics (such as CTL, LTL, ATL, epistemic logic, and their combinations) with semantics over modular interpreted systems.
By the same token, the model checking and satisfiability problems for those logics are well defined in MIS.

%% file: 22_mis_exmpl.tex
\section{Modeling with MIS}\label{sec:modeling}


We argue that the revised definition of MIS achieves a high level of
separation between components in a model.
The interaction between
an agent and the rest of the world is encapsulated in the agent's
interference functions $out_i,in_i$.
Of course, the design of the agent must take into account the tokens that can
be sent from modules with which the agent is supposed
to interact. For instance, the $out,in$ functions of two communicating agents
must be prepared to receive communication tokens from the other party.
However, the interference functions can be oblivious to the modules with which the agent does
not interact.
In this section, we demonstrate the advantages on two benchmark scenarios: Trains, Tunnel, and Controller (TTC), and Dining Cryptographers (DC).


%
%
\extended{
  \begin{remark}[Interaction vs. interference]
  A careful reader has probably noticed that we use the terms \emph{interaction}
  and \emph{interference} interchangeably.  This is because our representations
  should enable modeling of all kinds of interaction: not only conscious and
  purposeful (like communication), but also low-level interference that agents
  need neither to intended nor to be aware of.  For example, consider two robots
  whose bodies push each other without getting the other in the range of their
  sensors, two processes trying to print at the same time, players on the stock
  market and a joint-stock company. In all these scenarios, the current local
  state and local decision of one agent influences the transitions of other
  agents, usually without both agents being aware of that.

  Conscious and intended interaction is a special case of such general
  interference, in which the interference itself is represented in the local
  states of involved parties.
  \end{remark}
}

\newcommand{\itf}[1]{\mathsf{#1}}

\subsection{Tunnel, Trains, and Controller (TTC)}\label{sec:trains1}

TTC is a variant of classical mutual exclusion, and models $n$ trains
moving over cyclic tracks sharing a single tunnel.
Because only one train can be in the tunnel at a time, trains need to get a
permission from the controller before entering the tunnel.
We model the scenario by MIS $TTC_n = (Agt, Act, In)$, where:
\begin{itemize2}
\item $\Agtnames = \{ {tr}_1, \ldots, tr_n, \Controller \}$,
\item $\Act = \{ nop, approach, request, enter, leave \}$,
\item $\In = \{ \itf{idle}, \itf{appr}, \itf{try}_1, \ldots, \itf{try}_n,
  \itf{retry}, \itf{granted}, \itf{left}, \itf{enter}, \itf{aw\_reqs},
  \itf{grant}, \itf{grant}_1, \dots, \itf{grant}_n, \\ \itf{no\_reqs},
  \itf{infd}, \itf{ack\_release}, \itf{aw\_leave} \}$.
\item $\Agt = \{ \mathbf{tr}_1, \ldots, \mathbf{tr}_n, \mathbf{\Controller} \}$,
\end{itemize2}
The system includes $n$ trains
$\mathbf{tr}_i = \big(tr_i, \{(St_i, Init_i, d_i, out_i, in_i, o_i, \Pi_i, \pi_i)\}\big)$ for $i \in \{0,\ldots,n\}$ such that:

\smallskip
\begin{small}
\begin{tabular}{@{\!\!\!\!\!\!\!\!\!\!}p{0.5\textwidth}@{\quad\quad}p{0.5\textwidth}@{}}
  $St_i = \{ out, tun\_needed, granted, in \}$, \KolorAM{and $Init_i = \{ out
  \}$.}
  $d_i$ is defined as:
    \begin{itemize2}
      \item $out \mapsto \{ nop, approach \}$,
      \item $tun\_needed \mapsto \{ request \}$,
      \item $granted \mapsto \{ enter \}$,
      \item $in \mapsto \{ nop, leave \}$
    \end{itemize2}

  $o_i$ is defined as:
    \begin{itemize2}
      \item $((out, nop), \itf{idle}) \mapsto \{out\}$
      \item $((out, approach), \itf{appr})  \mapsto \{ tun\_needed \}$,
      \item $((tun\_needed, request), \itf{retry}) \mapsto \{ tun\_needed \}$,
      \item $((tun\_needed, request), \itf{granted}) \mapsto \{ granted \}$,
      \item $((granted, enter), \itf{enter}) \mapsto \{ in \}$,
      \item $((in, nop), \itf{idle}) \mapsto \{ in \}$,
      \item $((in, leave), \itf{leave})  \mapsto \{ out \}$
    \end{itemize2}

  $\Pi_i = \{ in\_tunnel \}$
%
&
%
  $out_i$ is defined as:
    \begin{itemize2}
      \item $(out, nop) \mapsto \{ \{ (\itf{idle}, tr_i) \} \}$,
      \item $(out, approach) \mapsto \{ \{ (\itf{appr}, tr_i) \} \}$,
      \item $(tun\_needed, request) \mapsto \{ \{ (\itf{try}_i, \Controller) \} \}$,
    \item $(granted, enter) \mapsto \{ \{ (\itf{enter}, tr_i) \} \}$,
    \item $(in, nop) \mapsto \{ \{ (\itf{idle}, tr_i) \} \}$,
    \item $(in, leave) \mapsto \{ \{ (\itf{left}, \Controller),
     (\itf{left},tr_i) \} \}$
    \end{itemize2}

  $in_i$ is defined as:
    \begin{itemize2}
      \item $s = out \land \itf{appr} \in H \mapsto \{ \itf{appr} \}$,
      \item $s = tun\_needed \land \itf{grant} \in H \mapsto \{ \itf{granted} \}$,
      \item $s = tun\_needed \mapsto \{ \itf{retry} \}$,
      \item $s = granted \land \itf{enter} \in H \mapsto \{ \itf{granted} \}$,
      \item $s = in \land \itf{left} \in H \mapsto \{ \itf{left} \}$,
      \item $\top \mapsto \{ \itf{idle} \}$
    \end{itemize2}

  $\pi_i = \{ in \mapsto in\_tunnel \}$
\end{tabular}
\end{small}

\vspace{0.1cm}
\noindent
Moreover, the agent
$\mathbf{\Controller} =
    \big(ctrl, \{(St_c, Init_c, d_c, out_c, in_c, o_c, \Pi_c, \pi_c)\}\big)$
modeling the controller is defined as follows:

\smallskip
\begin{small}
\begin{tabular}{@{\!\!\!\!\!\!\!\!\!\!}p{0.5\textwidth}@{\quad\quad}p{0.5\textwidth}@{}}
  $St_c = \{ tun\_free, infd, tr_{1}granted, \ldots, tr_{n}granted \}$,
  \KolorAM{and
    $Init_c = \{ tun\_free \}$.}

  $d_c$ is defined as:
    \begin{itemize2}
      \item $tun\_free \mapsto \{ accepting \}$,
      \item $infd \mapsto \{ waiting \}$,
      \item $tr_{1}granted \mapsto \{ inform \}$,
      \item \ldots
      \item $tr_{n}granted \mapsto \{ inform \}$
    \end{itemize2}

    $o_c$ is defined as:
        \begin{itemize2}
          \item \mbox{$((tun\_free, accepting), \itf{no\_reqs}) \mapsto$} $\{ tun\_free \}$,
          \item $((infd, waiting), \itf{aw\_leave}) \mapsto \{ infd \}$,
          \item $((infd, waiting), \itf{ack\_release}) \mapsto \{ tun\_free \}$,
          \item $((tun\_free, accepting), \itf{grant_1}) \mapsto \{ tr_{1}granted \}$,
          \item \ldots
          \item $((tun\_free, accepting), \itf{grant_n}) \mapsto \{ tr_{n}granted \}$,
          \item $((tr_{1}granted, inform), \itf{infd}) \mapsto \{ infd \}$,
          \item \ldots
          \item $((tr_{n}granted, inform), \itf{infd}) \mapsto \{ infd \}$
        \end{itemize2}
&
%
  $out_c$ is defined as:
    \begin{itemize2}
      \item $(tun\_free, accepting) \mapsto \{ \{ (\itf{aw\_reqs}, \epsilon) \} \}$,
      \item $(infd, waiting) \mapsto \{\{(\itf{aw\_leave}, \Controller)\} \}$,
      \item $(tr_{1}granted, inform) \mapsto \{\{(\itf{grant}, tr_1)\} \}$,
      \item \ldots
      \item $(tr_{n}granted, inform) \mapsto \{\{(\itf{grant}, tr_n)\} \}$,
    \end{itemize2}

  $in_c$ is defined as:
    \begin{itemize2}
      \item $s = tun\_free \land \itf{try}_1 \in H \mapsto \{
          \itf{grant_1} \}$,
        \item \ldots
        \item $s = tun\_free \land \itf{try}_n \in H \mapsto \{
            \itf{grant_n} \}$,
          \item $s = tun\_free \mapsto \{\itf{no\_reqs}\}$,
          \item $s = tr_{1}granted \lor ... \lor s = tr_{n}granted \mapsto
            \{ \itf{infd} \}$,
          \item $s = infd \land \itf{left} \in H \mapsto
            \{\itf{ack\_release}\}$,
          \item $s = infd \mapsto \{ \itf{aw\_leave} \}$,
          \item $\top \mapsto \{ \itf{idle} \}$
        \end{itemize2}

    \medskip
    $\Pi_c = \{ tunnel\_busy \}$

    $\pi_c = \{ infd \mapsto tunnel\_busy \}$
\end{tabular}
\end{small}

\begin{figure}[t]
  \centerline{\includegraphics[width=\textwidth]{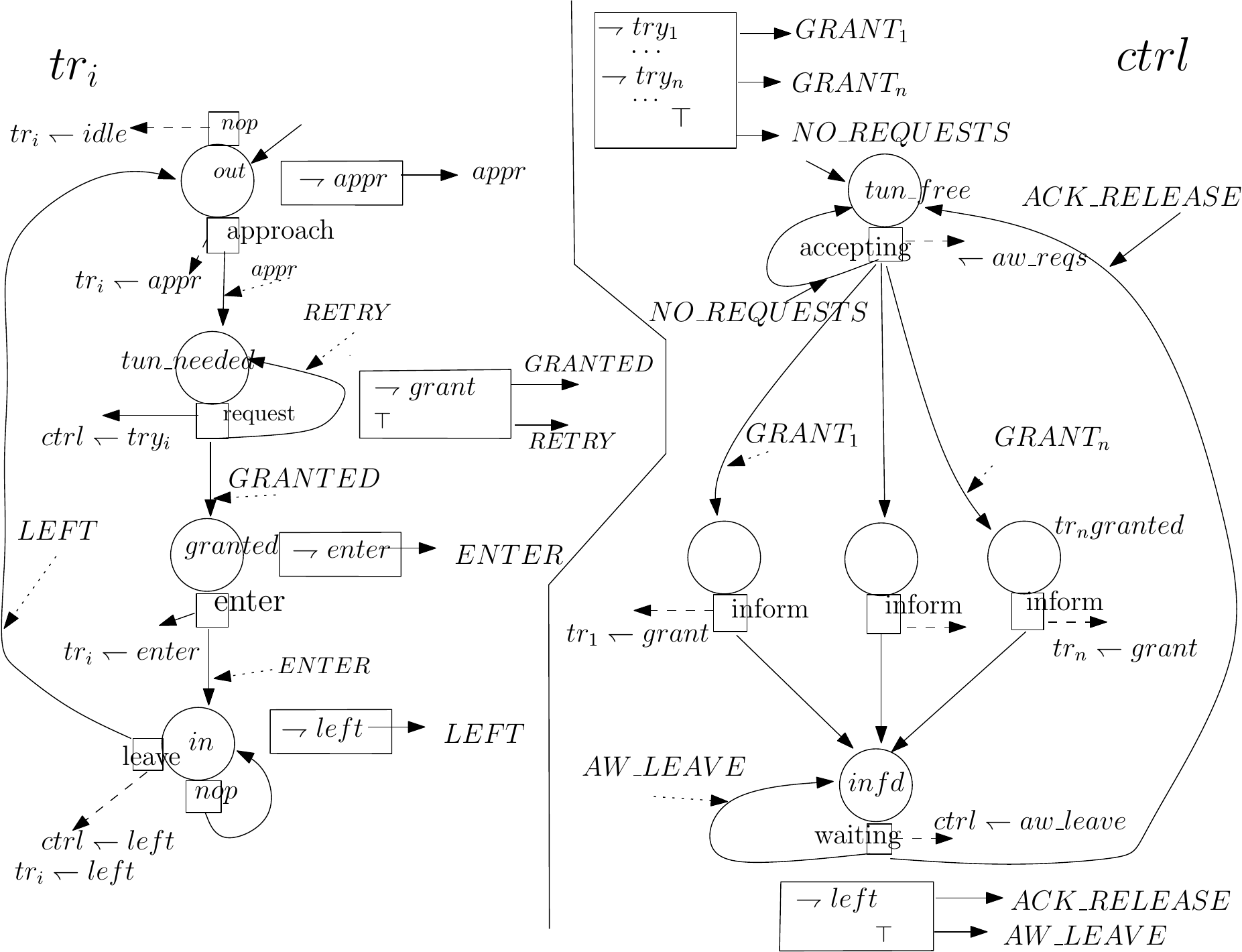}}
  \caption{Tunnel, trains and controller (TTC) in the graphical
  representation}
  \label{fig:trains1}
\end{figure}

The model is illustrated in Figure~\ref{fig:trains1} using the notation
introduced in Section~\ref{sec:graphical}. The protocol focuses on the procedure of
gaining a permission to access the tunnel. Before requesting the permission, a
train approaches the tunnel, and its state changes to $tun\_needed$. In this
state it requests the permission from the controller.  When the controller
grants the permission to one of the nondeterministically chosen trains
($tr_i{granted}$) it informs the train that got access to the tunnel about
this fact, and moves to the state $infd$. The train enters the tunnel in
the next step of the protocol, and changes its state to
$in$, whereas the remaining trains may continue requesting the access (they
remain in $tun\_needed$). When the train leaves the tunnel, it changes its
state to $tun\_free$.

\subsection{Graphical Representation}\label{sec:graphical}

As the definitions of MIS tend to be verbose, we introduce a simple graphical
notation, based on networks of communicating automata. Let us explain it,
based on Figure~\ref{fig:trains1}, which is a graphical representation of the
tunnel, trains, and controller model from Section~\ref{sec:trains1}:
%
%
\begin{itemize2}
  \item Modules  defining different {\bf agents} and belonging to the same agent are separated by solid and dashed lines, respectively,
  \item Circles correspond to {\bf local states}. An arrow with loose end pointing into a circle denotes an initial state,
  \item Boxes define {\bf local actions} associated with a state,
  \item For a local action, dashed lines going out of it define {\bf emitted influences}, specified with the receiver and the influence at the left and right side of an harpoon arrow pointed left, respectively. When no receiver is specified, the influence is broadcasted,
  \item Solid lines with arrows, connecting an action with a local state, correspond to a {\bf local transition function},
  \item For a local state, guarded commands (possibly in a box) define the translation of {\bf  external manifestations} received by
    an agent into {\bf local impressions}. A harpoon arrow pointed right corresponds to a sender at the left side and the message at the right side, and if the sender is not specified it means receiving from anyone.
    For a transition, dotted arrows pointing at it correspond to application of those impressions.
\end{itemize2}

The number of interactions $x$ received by an agent is denoted with $n(x)$.
The notation $*$ labeling a transition means that it is executed when none of
the remaining transitions are enabled. For example, it could be used instead
of directly specifying the generation and application of $aw\_leave$
manifestation in the controller.

Some parts can be skipped or abstracted away if it does not lead to confusion.
For example, if no influence is emitted and only one transition is associated
with an action, this action needs not be directly specified. In Figure
\ref{fig:trains1}, the self-loop from the $in$ state is not accompanied by the
associated local impression nor the impression.
Similarly, a single influence addressed to the very module that issued it can be omitted.
For example, we do not show the manifestation $idle$ in the graph.
Valuations of propositions can be depicted in a similar way as for networks of automata. We omit them in our examples throughout, as they do not play a role in this paper.

\begin{figure}[!t]
  \centerline{\includegraphics[width=\textwidth]{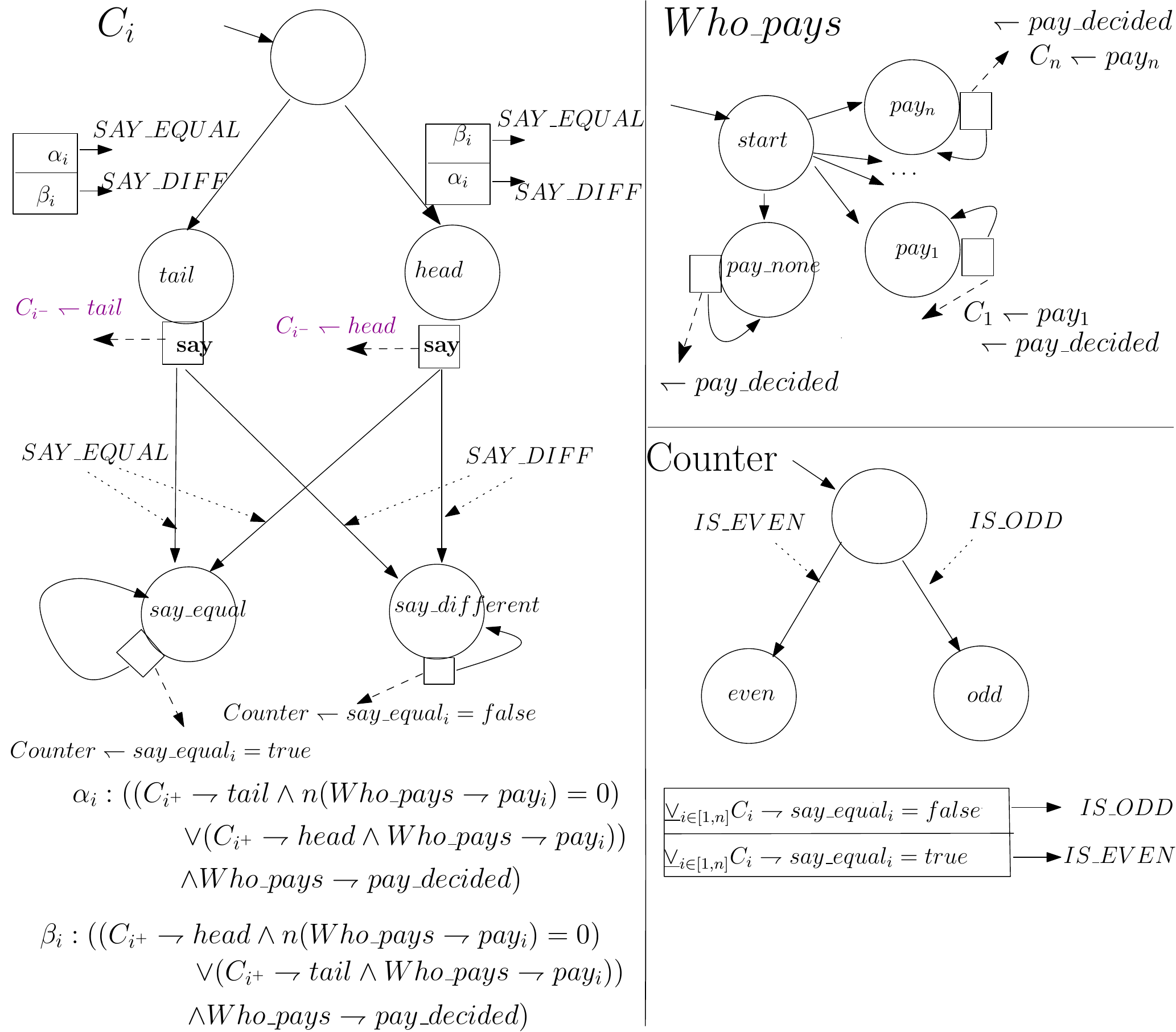}}
  \caption{MIS for dining cryptographers (DC1)}
  \label{fig:crypto-fixed}
\end{figure}

\subsection{Dining Cryptographers: Standard Version (DC1)}\label{sec:crypto-fixed}

Dining Cryptographers is a well-known benchmark proposed by
Chaum~\cite{Chaum}. $n$ cryptographers are having dinner, and the
bill is to be paid anonymously, either by one of them or by their employer.
In order to learn which option is the case without disclosing which cryptographer is paying (if any), they run a two-stage protocol.
First, every cryptographer is paired with precisely one other participant (they sit around the table), thus forming a cycle.
Every pair shares a one-bit secret, say by tossing a coin behind a menu.  In
the second stage, each cryptographer publicly announces whether he sees an odd or
an even number of coin heads, saying the opposite if being the payer.

In the simplest case (DC1) the number of cryptographers is fixed, and each
cryptographer is directly bound with its neighbours. Cryptographers announce
their utterances by broadcasting them.  A modular interpreted system modeling
this setting is presented in Figure~\ref{fig:crypto-fixed}.  For $n$
cryptographers, the $i$th cryptographer is modeled by agent $C_i$ ($0 \leq i
\le n$).  We introduce notation $i^+$ and $i^-$ to refer to the right and
the left neighbour of cryptographer $i$, respectively.
%
The system includes also two additional agents. $Who\_pays$ initializes the system by determining who is
the payer, and communicating it to the cryptographers. 
According to the protocol definition, either one of the cryptographers is
chosen, or none of the participants pays.  Agent $Counter$ counts the utterances of the cryptographers, computes the XOR operation (denoted by $\underline{\vee}$ and assuming that utterances {\em different} and {\em equal} correspond to true and false values, respectively, thus the result is true iff the number of {\em different} utterances is odd), and determines
the outcome of the protocol.  Figure~\ref{fig:crypto-fixed} shows the modular
interpreted system for DC1.

%% file: 30_agentivity.tex
\section{How to Measure Multi-Agency}\label{sec:modularity}

In this section, we present our preliminary attempt at defining what it means for a design to be multi-agent. Intuitively, separate agents should have only limited coordination and/or communication capabilities. Otherwise, the whole system can be seen as a single agent in disguise.
The idea is to measure the complexity of interference between different agents, and relate it to the complexity of the system. The former factor will be captured by the number of directed interaction tokens that a given agent can generate; the latter by the number of global transitions that can occur.
We say that the agent is well designed if its interference complexity is reasonably smaller than overall complexity of the system.

\begin{definition}[Interaction complexity]
The \emph{interaction complexity of agent $i$} in modular interpreted system $M$, denoted $IC(i)$, is defined as follows. Let $\#out_i(q_i)$ be the the maximal number of directed tokens generated by function $out_i$ to modules of other agents in state $q_i$. Furthermore, let $\#in_i(q_i)$ be the maximal number of tokens admitted by function $in_i$ from modules of other agents in state $q_i$. Now, $IC(i) = \sum_{q_i\in\States_i} (\#out_i(q_i) + \#in_i(q_i))$.

The \emph{interaction complexity of $M$} is defined as $IC(M) = \sum_{i\in\Agt} IC(i)$.
\end{definition}

\begin{definition}[Global complexity]
The \emph{global complexity of MIS $M$}, denoted $GC(M)$, is the number of transitions in the NCEGS unfolding of $M$.
\end{definition}

How can we express that $IC(M)$ is ``reasonably smaller'' than $GC(M)$?
Such a requirement is relatively easy to specify for \emph{classes} of models, parameterized with values of some parameter (for instance, the number of identical trains in the tunnel-controller scenario).

\begin{definition}[$\mathcal{C}$-sparse interaction, multi-agent design]
Let $\mathcal{M}$ be a class of MIS
and $\mathcal{C}$ a class of complexity functions $f : \complFn$.
We say that $\mathcal{M}$ is characterized by \emph{$\mathcal{C}$-sparse interaction} iff there is a function $f\in\mathcal{C}$ such that $IC(M) \le f(GC(M))$ for every $M\in \mathcal{M}$.

Furthermore, we say that $\mathcal{M}$ has \emph{multi-agent design} iff $\mathcal{M}$ has LOGTIME-sparse interaction, and $card(M)\ge 2$ for every $M\in\mathcal{M}$.
\end{definition}

\begin{proposition}
Classes TTC and DC1 have multi-agent design.
\end{proposition}
The proof is straightforward.
It is easy to see that the other variants of Dining Cryptographers, discussed in Section~\ref{sec:openness}, also have multi-agent design.

%% file: 40_openness.tex
\section{How Open is an Open System?}\label{sec:openness}

The idea of open systems is important for several communities: not only MAS, but also verification, software engineering, etc.
It is becoming even more important now, with modern technologies enabling dynamic networks of devices, users and services whose nodes can be created and removed according to current needs.
Traditionally, the term \emph{open system} is understood as a process coupled with the environment, which is rather disappointing given the highly distributed nature of MAS nowadays\short{.}
\extended{
  (think of mobile networks, wifi connections around a hotspot, users in a social network etc.).
}
One would rather like ``openness'' to mean that components (agents in our case) can freely join and leave the system without the need to redesign the rest of it.

Perfectly open systems are seldom in practice; usually, adding/removing components requires some transformation of the remaining part (for instance, if a server is to send personalized information to an arbitrary number of clients then it must add the name of each new client to the appropriate distribution lists).
So, it is rather the degree of openness that should be captured.
We try to answer the question \emph{How open is the system?} (or, to be more precise, its model) in the next subsection.

\subsection{A Measure of Openness}\label{sec:open-measure}

We base the measure on the following intuition: openness of a system is \emph{simplicity of adding and removing agents to and from the model}. That is, we consider two natural transformations of models: expansion (adding agents) and reduction (removing agents).
We note that the simplicity of a transformation is best measured by its algorithmic complexity, i.e., the number of steps needed to complete the transformation. A perfectly open system requires no transformation at all ($0$ steps) to accommodate new components, whereas at the other extreme we have systems that require redesigning of the model from scratch whenever a new agent arrives.

Note that the openness of a model depends on \emph{which} agents want to join or leave. For instance, the system with trains and controllers should be able to easily accommodate additional trains, but not necessarily additional controllers. Likewise, departure of a train should be straightforward, but not necessarily that of the controller.
No less importantly, the context matters. We are usually not interested in an arbitrary expansion or reduction (which are obviously trivial). We want to add or remove agents while keeping the ``essence'' of the system's behavior intact.
The following definitions formalize the idea.

\begin{definition}[Expansion and reduction of a MIS]
Let $M = (\Agtnames,\Act,\In,\Agt)$ be a MIS, and $\agent$ an agent (in the sense of Definition~\ref{def:new-mis}).
By $agt(\agent)$ (resp. $act(\agent)$, $in(\agent)$) we denote the set of agent names (resp. action symbols, interaction symbols) occurring in $\agent$.
Moreover, $ns(\agent,M)$ will denote the set of $\agent$'s namesakes in $M$.\footnote{
  That is, agents in $M$ that have the same id as $\agent$. }
Note that $ns(\agent,M)$ can contain at most $1$ agent.

The \emph{expansion of $M$ by $\agent$} is defined as the modular interpreted system $M\oplus\agent = (\Agtnames',\Act',\In',\Agt')$ where:\ $\Agtnames'=\Agtnames\cup agt(\agent)$,\ $\Act'=\Act\cup act(\agent)$,\ $\In'=\In\cup in(\agent)$, and\ $\Agt'=\Agt\setminus ns(\agent,M)\cup\{\agent\}$.
\
The \emph{reduction of $M$ by $\agent$} is defined as $M\ominus\agent = (\Agtnames,\Act,\In,\Agt')$ where $\Agt'=\Agt\setminus\{\agent\}$.
\end{definition}

Thus, expansion corresponds to ``dumb'' pasting an agent into a MIS, and reduction corresponds to simple removal of the agent. The operations are well defined in the following sense.
\begin{proposition}
Expansion/reduction of a MIS is always a MIS.\footnote{
  The proofs of results in Section~\ref{sec:openness} are straightforward from the construction of MIS, and we leave them to the reader. }
\end{proposition}
\extended{
  \begin{myproof}
  It follows by the construction of MIS. \noteWJ{TO BE COMPLETED (.... what about directed manifestations?????)}
  \end{myproof}
}

It is easy to see that removing an agent and pasting it in again does not change the MIS. The reverse sequence of operations does change the MIS. However, both structures have the same unfoldings:
\begin{proposition}
Let $\agent$ be an agent in $M$. Then, $(M\ominus\agent)\oplus\agent = M$.
\
Moreover, let $\agent$ be an agent with no namesake in $M$. Then, $NCEGS((M\oplus\agent)\ominus\agent) = NCEGS(M)$.
\end{proposition}
\extended{
  \begin{myproof}
  \noteWJ{TO BE COMPLETED ???It follows by the construction of MIS. (because unused symbols do not matter)}
  \end{myproof}
}

Now we can make our first attempt at a measure of openness.
\begin{definition}[Degree of openness]
Let $\theta$ be a property of models,\footnote{
  We do not restrict the language in which $\theta$ is specified. It can be propositional logic, first-order temporal logic, or even the general language of mathematics. The only requirement is that, for every MIS $M$, the truth of $\theta$ in $M$ is well defined. }
$M$ a modular interpreted system, and $\agent$ an agent.
The \emph{degree of openness} of $M$ wrt expansion by $\agent$ under constraint $\theta$ is defined as the minimal number of steps that transform $M\oplus\agent$ into a MIS $M'$ such that $card(M')=card(M\oplus\agent)$ and $M'$ satisfies $\theta$.

Likewise, the degree of openness of modular interpreted system $M$ wrt reduction by agent $\agent$ under constraint $\theta$ is the minimal number of steps that transform $M\ominus\agent$ into an $M'$ such that $card(M')=card(M\ominus\agent)$ and $M'$ satisfies $\theta$.
\end{definition}

The constraint $\theta$ can for example refer to liveness of the system or some of its components, fairness in access to some resources, and/or safety of critical sections.
Note that the cardinality check is essential in the definition -- otherwise, a possible transformation would be to simply delete the newly added agent from $M\oplus\agent$ (respectively, to restore \agent in $M\ominus\agent$).

\begin{definition}[Openness of a class of models]
Let $\mathcal{M}$ be a class of MIS, $\agent$ an agent, and $\theta$ a property of models.
Moreover, let $\mathcal{C}$ be a class of complexity functions $f : \complFn$.
$\mathcal{M}$ is \emph{$\mathcal{C}$-open} wrt expansion (resp. reduction) by $\agent$ under constraint $\theta$ iff there is a complexity function $f\in\mathcal{C}$ such that for every $M\in\mathcal{M}$ the degree of openness of $M$ wrt expansion (resp. reduction) by $\agent$ under $\theta$ is no greater than  $f(|M|)$.
\end{definition}

The most cumbersome part of the above definitions is the constraint $\theta$.
How can one capture the ``essence'' of acceptable expansions and reductions?
Note that, semantically, $\theta$ can be seen as a subclass of models.
We postulate that in most scenarios the class that defines acceptable expansions/reductions is the very class whose openness we want to measure. This leads to the following refinement of the previous definitions.

\begin{definition}[Openness in a class]
The degree of openness of $M$ wrt expansion (resp. reduction) by $\agent$ in class $\mathcal{M}$ is the minimal number of steps that transform $M\oplus\agent$ (resp. $M\ominus\agent$) into a MIS $M'\in\mathcal{M}$ such that $card(M')=card(M\oplus\agent)$.

Moreover, $\mathcal{M}$ is $\mathcal{C}$-open wrt expansion (resp. reduction) by $\agent$ iff there is a complexity function $f\in\mathcal{C}$ such that for every $M\in\mathcal{M}$ the degree of openness of $M$ wrt expansion (resp. reduction) by $\agent$ in $\mathcal{M}$ is no greater than $f(|M|)$.
\end{definition}

We explain the measure in greater detail in the remainder of Section~\ref{sec:openness}. It is important to note that (in contrast to the measure of multi-agentivity proposed in Section~\ref{sec:modularity}) our measure of openness is not specific to MIS, and can be applied to other modeling frameworks.

\begin{remark}
Alternatively, we could define the openness of $M$ wrt $\agent$ and $\theta$ by the \emph{Kolmogorov complexity} of an appropriate expansion/reduction, i.e., by the size of the shortest algorithm that transforms $M$ in an appropriate way.
We chose time complexity instead, for two reasons.
First, Kolmogorov complexity often obscures the level of difficulty of a process (e.g., a two-line algorithm with an infinite \textbf{while} loop can implement infinitely many changes, which gives the same complexity as changing the names of two communication channels for a controller). Secondly, computing Kolmogorov complexity can be cumbersome as it is Turing-equivalent to answering the halting problem.

We observe, however, that a Kolmogorov-style measure of openness can be a good alternative for infinite models, especially ones that require infinitely many steps to accommodate changes in the configuration of components.
\end{remark}

\extended{
  Below, we apply the measure of openness to the example presented in
  Section~\ref{sec:trains1}.
  \begin{example}
      In the case of Tunnel, Trains, and Controller example, we can assume that
      the property~$\theta$ describes the fact that the mutual exclusion
      property is preserved, and the communication with the controller is
      implemented in the same coherent way for all the agents in the system.
      Then, we can estimate
      the size of the representation of our system consisting of $n$ trains to
      be equal $106n+37$ symbols. Therefore, the number of new symbols that need adding
      (removing) to extend (reduce) our system by one train, equals~$106$.
      However, when adding $n$ new trains, only $19n$ symbols of the controller
      module need to be added (removed) assuming that we count only the modifications of
      the agents that are not being added (removed), and this gives us the
      openness of TTC.
  \end{example}

}

\subsection{How to Open Up Cryptographers}

In Section \ref{sec:crypto-fixed} we modeled the standard version of the Dining Cryptographers protocol as a modular interpreted system (class DC1). In this section, we will determine the openness of DC1, plus two other classes of MIS modeling other versions of the protocol. To comply with classical rules of composition, we begin with the least open variant.

\begin{figure}[!t]
\centerline{\includegraphics[width=\textwidth]{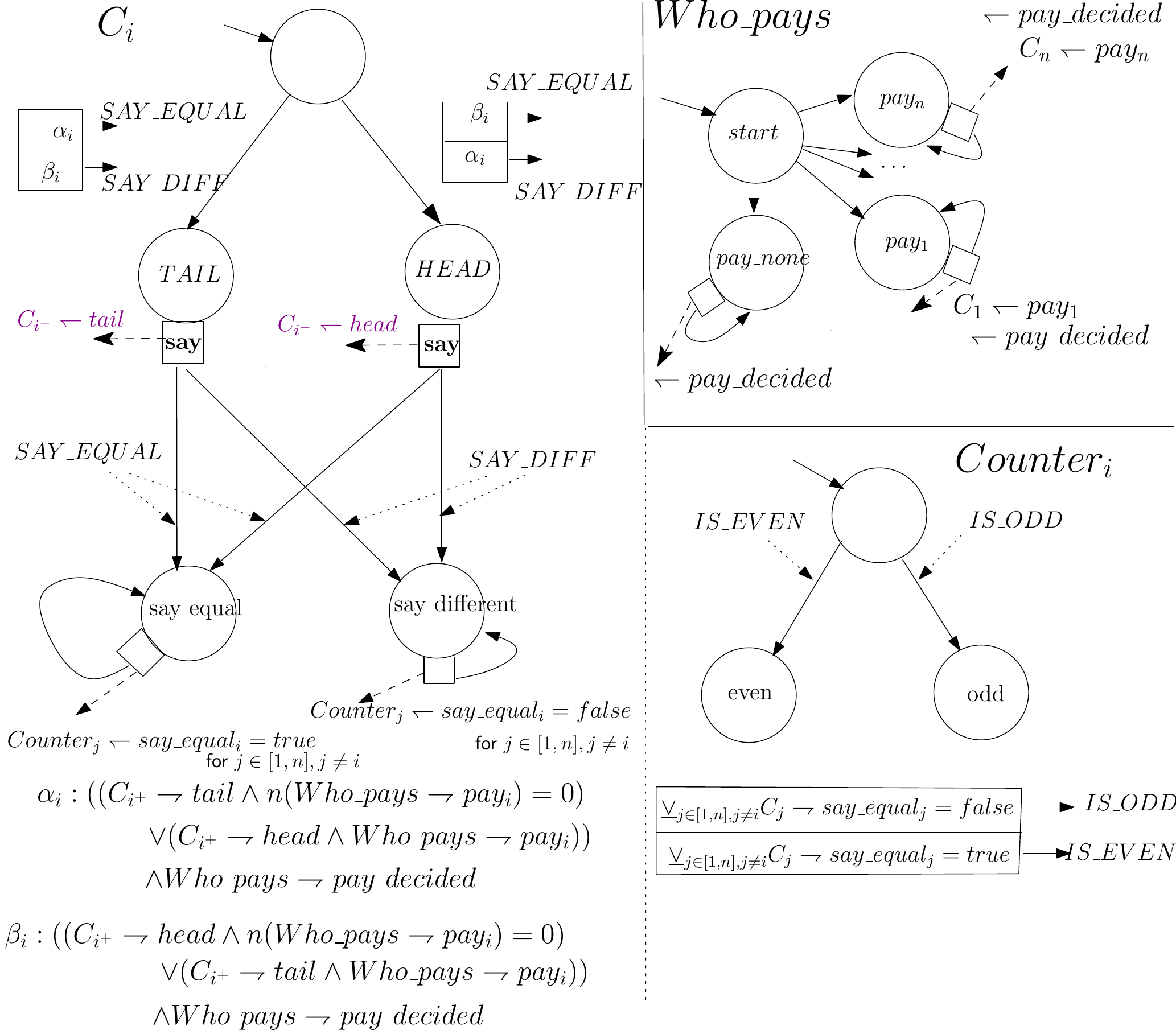}}
\caption{Dining cryptographers version DC2: direct channels instead of broadcast}
\label{fig:crypto-DC2}
\end{figure}

\subsubsection{DC-Net, Direct Channels, No Broadcasting (DC2)}

Let us assume that no broadcast channel is available, or it is too faulty (or insecure) to be of use in multi-party computation. In such case, every pair of cryptographers must use a direct secured channel for communicating the final utterance.
The result of the computation is calculated independently by every cryptographer.
We denote this class of models by DC2, and construct it as follows.
Each cryptographer $i$ is modeled by agent $C_i$, similar to the cryptographer agents in DC1. Instead of a single global counter of utterances, there is one counter per every cryptographer ($Counter_i$). The final utterance is sent by direct point-to-point channels to the counters of all other participants.
The resulting MIS is shown in Figure \ref{fig:crypto-DC2}.

Adding a new cryptographer $C_i$ to DC1$_n$ requires the following changes.
First, modifying links among the new neighbours of $C_i$ yields $10$ changes. Secondly,
every agent $C_j$ in DC1$_n$ must be modified in order to establish a communication channel with $C_i$.
This requires $2\cdot 5$ changes per cryptographer, thus $10n$ changes are needed.
Thirdly, for the agent $Who\_pays$, we add the state $pay_{i'}$ with corresponding transitions:
a single non-deterministic transition from $start$ to $pay_{i'}$ (17 steps:
$2$ for $d_i$ + $4$ for $out_i$ + $8$ for $in_i$ + $3$ for $o_i$), and the loop sending payment information (19 steps: $4$ for $d_i$ + $4$ for $out_i$ + $4$ for $in_i$ + $3$ for $o_i$).
Finally, $Counter$ needs to be updated to take into account the new participant. A XOR argument is added with  receiving a manifestation, yielding $2\cdot 4 = 8$ changes.
Thus, the overall openness complexity for DC2$_n$ is $10n + 54$.

\begin{proposition}\label{prop:DC2}
Class DC2 is $O(n)$-open wrt expansion by a cryptographer.
\end{proposition}

\subsubsection{Dining Cryptographers: Standard Version with Broadcast (DC1)}\label{sec:crypto-fixed-open}



Let us now go back to the standard version of the protocol, presented in Section~\ref{sec:crypto-fixed}
Adding a new cryptographer $C_i$ requires the following changes. First, modifying links for the new neighbors of $C_i$ requires 10 changes. Secondly, changes in $Who\_pays$ and $Counter$ are the same as for DC2$_n$, yielding $44$ steps.
Thus, $54$ changes are needed to accommodate the new cryptographer, regardless of the number of agents already present in the system.

\begin{proposition}\label{prop:DC1}
Class DC1 is $O(1)$-open wrt expansion by a cryptographer.
\end{proposition}

\begin{figure}[!t]
\centerline{\includegraphics[width=0.91\textwidth]{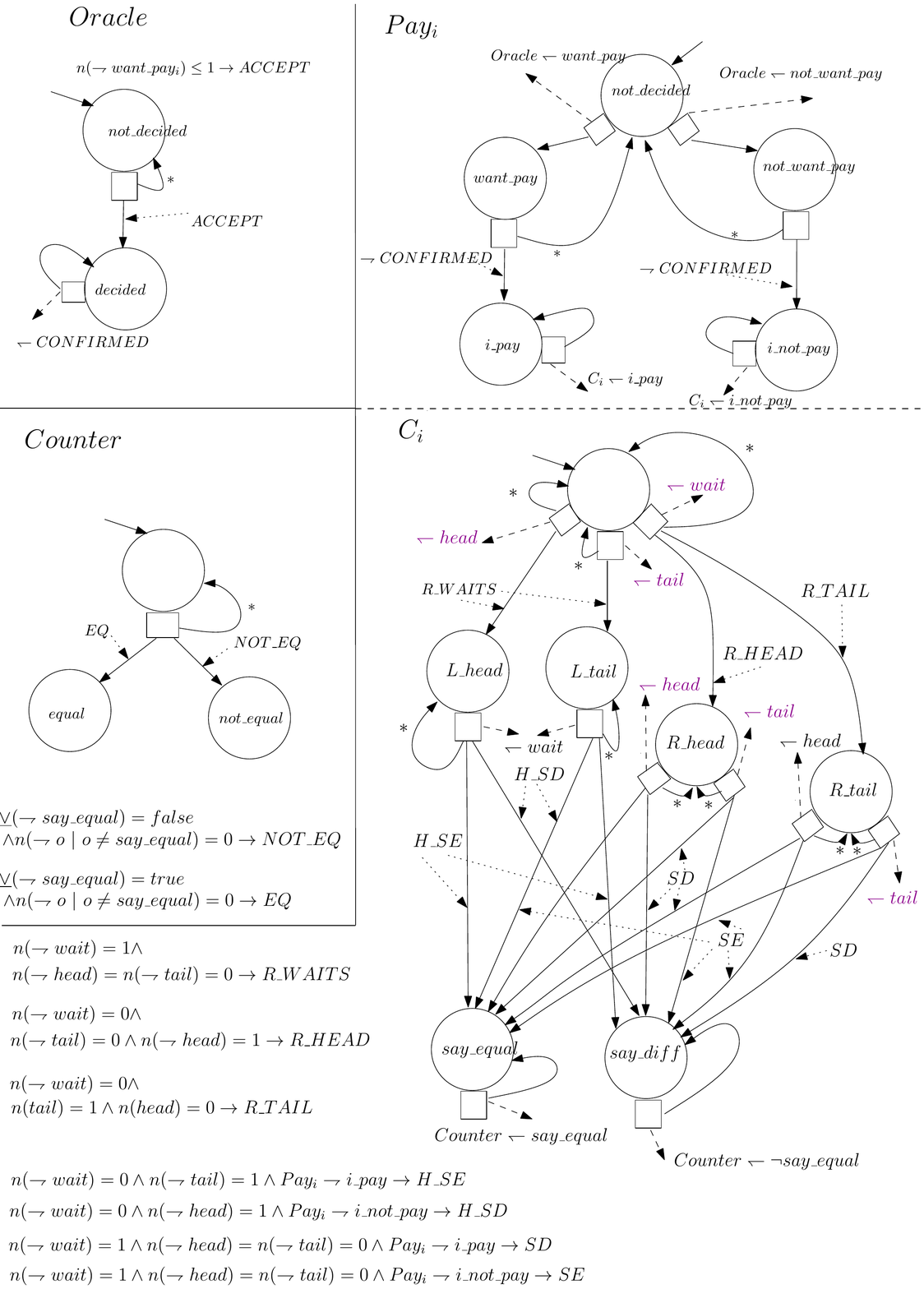}}
\caption{Cryptographers without identifiers (DC0)}
\label{fig:crypto-noid}
\end{figure}

\subsubsection{Fully Open System, Cryptographers without Identifiers (DC0)}

In our most radical variant, cryptographers are not arbitrarily assigned as neighbors. Instead, they  establish their neighborhood relation on their own before starting the protocol. Every cryptographer is modeled by two modules $C_i$ and $Pay_i$, and there are two additional agents $Oracle$ and $Counter$, cf.~Figure~\ref{fig:crypto-noid}. The system proceeds as follows:
\begin{description}
\item[Setting up the payer.] Every cryptographer sends the oracle his declaration whether he is going to pay or not (chosen nondeterministically). This is performed by module $Pay_i$. If $Oracle$ receives at most one statement $want\_pay$, it confirms to all cryptographers. If more  than one statements $want\_pay$ is sent, the round is repeated until the payment issue becomes resolved.

 \item[Establishing the neighbourhood relation and tossing coins.] Each cryptographer either nondeterministically tosses a coin and announces the outcome, or listens to such announcements from the other agents.
     If there is exactly one cryptographer announcing and one listening, they become paired. They register the value of the announcement, and proceed further. A cryptographer who started with announcing will now listen, and vice versa. This takes several rounds, and completes when every cryptographer has been paired with two neighbors (one to whom he listened, and one to whom he announced).

\item[Computation.] A broadcast channel is used for sending around the utterances ($say\_equal$ or $\neg say\_equal$). $Counter$ counts the utterances and computes their XOR on the spot, in the way described before.
\end{description}

\medskip
DC0 is fully open, as adding a new cryptographer requires no adaptation of DC0$_n$.

\begin{proposition}\label{prop:DC0}
Class DC0 is $O(0)$-open wrt expansion by a cryptographer.
\end{proposition}

By comparing their classes of openness, it is clear that DC1 is significantly more open wrt expansion than DC2 (constant vs. linear openness). On the other hand, it seems that the gap between DC1 and DC0 is rather slight ($O(1)$ vs. $O(0)$). Is that really the case? We believe that the difference between $O(1)$-openness and $O(0)$-openness is larger than one is used to in complexity of algorithms. First, constant openness means that, when expanding the MIS by a \emph{set} of new agents, the required transformation can be linear in the size of the set. More importantly, non-zero openness signifies the need to come up with a \emph{correct} procedure of expansion. In contrast, zero openness means zero hassle: the new agents can join the system as they come. There is no need for ``maintenance'' of the system so that it stays compliant with its (usually implicit) specification.

%% file: 50_conclusions.tex
\section{Conclusions}\label{sec:conclusions}

In this paper, we propose a new version of modular interpreted systems. The aim is to let modeling and analysis of multi-agent systems benefit from true separation of interference between agents and the ``internals'' of their processes that go on in a system. Thanks to that, one can strive for a more modular and open design. Even more importantly, one can use the MIS representation of a system to assess its agentivity and openness through application of simple mathematical measures.

We emphasize that it was \emph{not} our aim to create yet another agent programming language or representations that will be used as input to cutting-edge model checkers. Instead, we propose a class of models which enables to expose the internal structure of a multi-agent system, and to
define the concepts of openness and multi-agentivity in a precise mathematical sense.
While our definition of multi-agentivity is specific to MIS, the measure of openness is in fact generic, and can be applied to models defined in other formalisms (such as Reactive Modules). We plan to look closer at the degree of openness provided by different representation frameworks in the future.

We would also like to stress that the focus of this paper regarding the measures of agentivity and openness is on formalizing the concepts and showing how they work on benchmarks.
An formal study of the measures and their properties is a matter of future work.

\medskip\noindent\textbf{Acknowledgements.} The authors thank Andrzej Tarlecki for his suggestion to improve modularity of MIS by using multisets, and Thomas Agotnes for discussions.
Wojciech Jamroga acknowledges the support of the FNR (National Research Fund) Luxembourg under project GALOT -- INTER/DFG/12/06.
Artur \Meski acknowledges the support of the European Union, European Social Fund. Project PO KL
``Information technologies: Research and their interdisciplinary
applications'' (UDA-POKL.04.01.01-00-051/10-00).

%% file: mis.bbl
\begin{thebibliography}{10}
\providecommand{\bibitemdeclare}[2]{}
\providecommand{\surnamestart}{}
\providecommand{\surnameend}{}
\providecommand{\urlprefix}{Available at }
\providecommand{\url}[1]{\texttt{#1}}
\providecommand{\href}[2]{\texttt{#2}}
\providecommand{\urlalt}[2]{\href{#1}{#2}}
\providecommand{\doi}[1]{doi:\urlalt{http://dx.doi.org/#1}{#1}}
\providecommand{\bibinfo}[2]{#2}

\bibitemdeclare{article}{DBLP:journals/fmsd/AlurH99b}
\bibitem{DBLP:journals/fmsd/AlurH99b}
\bibinfo{author}{R.~\surnamestart Alur\surnameend} \& \bibinfo{author}{T.~A.
  \surnamestart Henzinger\surnameend} (\bibinfo{year}{1999}):
  \emph{\bibinfo{title}{Reactive Modules}}.
\newblock {\sl \bibinfo{journal}{Formal Methods in System Design}}
  \bibinfo{volume}{15}(\bibinfo{number}{1}), pp. \bibinfo{pages}{7--48},
  \doi{10.1023/A:1008739929481}.

\bibitemdeclare{article}{Alur02ATL}
\bibitem{Alur02ATL}
\bibinfo{author}{R.~\surnamestart Alur\surnameend}, \bibinfo{author}{T.~A.
  \surnamestart Henzinger\surnameend} \& \bibinfo{author}{O.~\surnamestart
  Kupferman\surnameend} (\bibinfo{year}{2002}):
  \emph{\bibinfo{title}{{A}lternating-Time {T}emporal {L}ogic}}.
\newblock {\sl \bibinfo{journal}{Journal of the ACM}} \bibinfo{volume}{49}, pp.
  \bibinfo{pages}{672--713}, \doi{10.1145/585265.585270}.

\bibitemdeclare{inproceedings}{Alur99communicatingHSM}
\bibitem{Alur99communicatingHSM}
\bibinfo{author}{R.~\surnamestart Alur\surnameend},
  \bibinfo{author}{S.~\surnamestart Kannan\surnameend} \&
  \bibinfo{author}{M.~\surnamestart Yannakakis\surnameend}
  (\bibinfo{year}{1999}): \emph{\bibinfo{title}{Communicating Hierarchical
  State Machines}}.
\newblock In: {\sl \bibinfo{booktitle}{Proceedings of ICALP}}, pp.
  \bibinfo{pages}{169--178}, \doi{10.1007/3-540-48523-6\_14}.

\bibitemdeclare{phdthesis}{Calta12phd}
\bibitem{Calta12phd}
\bibinfo{author}{J.~\surnamestart Calta\surnameend} (\bibinfo{year}{2012}):
  \emph{\bibinfo{title}{Synthesis of Strategies for Multi-Agent Systems}}.
\newblock Ph.D. thesis, \bibinfo{school}{Humboldt University Berlin}.

\bibitemdeclare{article}{Chaum}
\bibitem{Chaum}
\bibinfo{author}{D.~\surnamestart Chaum\surnameend} (\bibinfo{year}{1988}):
  \emph{\bibinfo{title}{The Dining Cryptographers Problem: Unconditional Sender
  and Recipient Untraceability}}.
\newblock {\sl \bibinfo{journal}{Journal of Cryptology}}
  \bibinfo{volume}{1(1)}, pp. \bibinfo{pages}{65--75},
  \doi{10.1007/BF00206326}.

\bibitemdeclare{incollection}{Dembinski03verics}
\bibitem{Dembinski03verics}
\bibinfo{author}{P.~\surnamestart Dembi\'nski\surnameend},
  \bibinfo{author}{A.~\surnamestart Janowska\surnameend},
  \bibinfo{author}{P.~\surnamestart Janowski\surnameend},
  \bibinfo{author}{W.~\surnamestart Penczek\surnameend},
  \bibinfo{author}{A.~\surnamestart P\'o{\l}rola\surnameend},
  \bibinfo{author}{M.~\surnamestart Szreter\surnameend},
  \bibinfo{author}{B.~\surnamestart Wo\'zna\surnameend} \&
  \bibinfo{author}{A.~\surnamestart Zbrzezny\surnameend}
  (\bibinfo{year}{2003}): \emph{\bibinfo{title}{Verics: A Tool for Verifying
  Timed Automata and Estelle Specifications}}.
\newblock In: {\sl \bibinfo{booktitle}{Proceedings of the of the 9th Int. Conf.
  on Tools and Algorithms for Construction and Analysis of Systems
  (TACAS'03)}}, {\sl \bibinfo{series}{LNCS}} \bibinfo{volume}{2619},
  \bibinfo{publisher}{Springer}, pp. \bibinfo{pages}{278--283},
  \doi{10.1007/3-540-36577-X\_20}.

\bibitemdeclare{incollection}{Emerson90temporal}
\bibitem{Emerson90temporal}
\bibinfo{author}{E.~A. \surnamestart Emerson\surnameend}
  (\bibinfo{year}{1990}): \emph{\bibinfo{title}{Temporal and Modal Logic}}.
\newblock In \bibinfo{editor}{J.~\surnamestart van Leeuwen\surnameend}, editor:
  {\sl \bibinfo{booktitle}{Handbook of Theoretical Computer Science}},
  \bibinfo{volume}{B}, \bibinfo{publisher}{Elsevier Science Publishers}, pp.
  \bibinfo{pages}{995--1072}.

\bibitemdeclare{book}{Fagin95knowledge}
\bibitem{Fagin95knowledge}
\bibinfo{author}{R.~\surnamestart Fagin\surnameend}, \bibinfo{author}{J.~Y.
  \surnamestart Halpern\surnameend}, \bibinfo{author}{Y.~\surnamestart
  Moses\surnameend} \& \bibinfo{author}{M.~Y. \surnamestart Vardi\surnameend}
  (\bibinfo{year}{1995}): \emph{\bibinfo{title}{Reasoning about Knowledge}}.
\newblock \bibinfo{publisher}{MIT Press}.

\bibitemdeclare{inproceedings}{DBLP:conf/concur/FisherHNPSV11}
\bibitem{DBLP:conf/concur/FisherHNPSV11}
\bibinfo{author}{J.~\surnamestart Fisher\surnameend}, \bibinfo{author}{T.~A.
  \surnamestart Henzinger\surnameend}, \bibinfo{author}{D.~\surnamestart
  Nickovic\surnameend}, \bibinfo{author}{N.~\surnamestart Piterman\surnameend},
  \bibinfo{author}{A.~V. \surnamestart Singh\surnameend} \&
  \bibinfo{author}{M.~Y. \surnamestart Vardi\surnameend}
  (\bibinfo{year}{2011}): \emph{\bibinfo{title}{Dynamic Reactive Modules}}.
\newblock In: {\sl \bibinfo{booktitle}{Proceedings of CONCUR}}, pp.
  \bibinfo{pages}{404--418}, \doi{10.1007/978-3-642-23217-6\_27}.

\bibitemdeclare{book}{Gecseg86automata}
\bibitem{Gecseg86automata}
\bibinfo{author}{F.~\surnamestart Gecseg\surnameend} (\bibinfo{year}{1986}):
  \emph{\bibinfo{title}{Products of Automata}}.
\newblock \bibinfo{series}{EATCS Monographs on Theor. Comput. Sci.},
  \bibinfo{publisher}{Springer}, \doi{10.1007/978-3-642-61611-2}.

\bibitemdeclare{article}{Holzmannn97spin}
\bibitem{Holzmannn97spin}
\bibinfo{author}{G.~J. \surnamestart Holzmannn\surnameend}
  (\bibinfo{year}{1997}): \emph{\bibinfo{title}{The Model Checker {SPIN}}}.
\newblock {\sl \bibinfo{journal}{IEEE Transactions on Software Engineering}}
  \bibinfo{volume}{23}(\bibinfo{number}{5}), pp. \bibinfo{pages}{279--295},
  \doi{10.1109/32.588521}.

\bibitemdeclare{techreport}{Jamroga06mis-tr}
\bibitem{Jamroga06mis-tr}
\bibinfo{author}{W.~\surnamestart Jamroga\surnameend} \&
  \bibinfo{author}{T.~\surnamestart {\AA}gotnes\surnameend}
  (\bibinfo{year}{2006}): \emph{\bibinfo{title}{Modular Interpreted Systems: A
  Preliminary Report}}.
\newblock \bibinfo{type}{Technical Report} \bibinfo{number}{IfI-06-15},
  \bibinfo{institution}{Clausthal University of Technology}.

\bibitemdeclare{inproceedings}{Jamroga07mis-aamas}
\bibitem{Jamroga07mis-aamas}
\bibinfo{author}{W.~\surnamestart Jamroga\surnameend} \&
  \bibinfo{author}{T.~\surnamestart {\AA}gotnes\surnameend}
  (\bibinfo{year}{2007}): \emph{\bibinfo{title}{Modular Interpreted Systems}}.
\newblock In: {\sl \bibinfo{booktitle}{Proceedings of AAMAS'07}}, pp.
  \bibinfo{pages}{892--899}, \doi{10.1145/1329125.1329286}.

\bibitemdeclare{inproceedings}{Koester11abstraction}
\bibitem{Koester11abstraction}
\bibinfo{author}{M.~\surnamestart K{\"o}ster\surnameend} \&
  \bibinfo{author}{P.~\surnamestart Lohmann\surnameend} (\bibinfo{year}{2011}):
  \emph{\bibinfo{title}{Abstraction for model checking modular interpreted
  systems over {ATL}}}.
\newblock In: {\sl \bibinfo{booktitle}{Proceedings of AAMAS}}, pp.
  \bibinfo{pages}{1129--1130}.

\bibitemdeclare{article}{Laroussinie08expATL}
\bibitem{Laroussinie08expATL}
\bibinfo{author}{F.~\surnamestart Laroussinie\surnameend},
  \bibinfo{author}{N.~\surnamestart Markey\surnameend} \&
  \bibinfo{author}{G.~\surnamestart Oreiby\surnameend} (\bibinfo{year}{2008}):
  \emph{\bibinfo{title}{On the Expressiveness and Complexity of {ATL}}}.
\newblock {\sl \bibinfo{journal}{Logical Methods in Computer Science}}
  \bibinfo{volume}{4}, p.~\bibinfo{pages}{7}, \doi{10.2168/LMCS-4(2:7)2008}.

\bibitemdeclare{inproceedings}{Lichtenstein85checking}
\bibitem{Lichtenstein85checking}
\bibinfo{author}{O.~\surnamestart Lichtenstein\surnameend} \&
  \bibinfo{author}{A.~\surnamestart Pnueli\surnameend} (\bibinfo{year}{1985}):
  \emph{\bibinfo{title}{Checking that finite state concurrent programs satisfy
  their linear specification}}.
\newblock In: {\sl \bibinfo{booktitle}{POPL '85: Proceedings of the 12th ACM
  SIGACT-SIGPLAN symposium on Principles of programming languages}},
  \bibinfo{publisher}{ACM}, \bibinfo{address}{New York, NY, USA}, pp.
  \bibinfo{pages}{97--107}, \doi{10.1145/318593.318622}.

\bibitemdeclare{inproceedings}{Lomuscio06mcmas}
\bibitem{Lomuscio06mcmas}
\bibinfo{author}{A.~\surnamestart Lomuscio\surnameend} \&
  \bibinfo{author}{F.~\surnamestart Raimondi\surnameend}
  (\bibinfo{year}{2006}): \emph{\bibinfo{title}{{MCMAS} : A Model Checker for
  Multi-agent Systems}}.
\newblock In: {\sl \bibinfo{booktitle}{Proceedings of TACAS}}, {\sl
  \bibinfo{series}{Lecture Notes in Computer Science}} \bibinfo{volume}{4314},
  pp. \bibinfo{pages}{450--454}, \doi{10.1007/11691372\_31}.

\bibitemdeclare{inproceedings}{Murano08hierarchical}
\bibitem{Murano08hierarchical}
\bibinfo{author}{A.~\surnamestart Murano\surnameend},
  \bibinfo{author}{M.~\surnamestart Napoli\surnameend} \&
  \bibinfo{author}{M.~\surnamestart Parente\surnameend} (\bibinfo{year}{2008}):
  \emph{\bibinfo{title}{Program Complexity in Hierarchical Module Checking}}.
\newblock In: {\sl \bibinfo{booktitle}{Proceedings of LPAR}}, pp.
  \bibinfo{pages}{318--332}, \doi{10.1007/978-3-540-89439-1\_23}.

\bibitemdeclare{phdthesis}{Raimondi06phd}
\bibitem{Raimondi06phd}
\bibinfo{author}{F.~\surnamestart Raimondi\surnameend} (\bibinfo{year}{2006}):
  \emph{\bibinfo{title}{Model Checking Multi-Agent Systems}}.
\newblock Ph.D. thesis, \bibinfo{school}{University College London}.

\bibitemdeclare{article}{Torre08HSM}
\bibitem{Torre08HSM}
\bibinfo{author}{S.~La \surnamestart Torre\surnameend},
  \bibinfo{author}{M.~\surnamestart Napoli\surnameend},
  \bibinfo{author}{M.~\surnamestart Parente\surnameend} \&
  \bibinfo{author}{G.~\surnamestart Parlato\surnameend} (\bibinfo{year}{2008}):
  \emph{\bibinfo{title}{Verification of scope-dependent hierarchical state
  machines}}.
\newblock {\sl \bibinfo{journal}{Information and Computation}}
  \bibinfo{volume}{206}(\bibinfo{number}{9-10}), pp.
  \bibinfo{pages}{1161--1177}, \doi{10.1016/j.ic.2008.03.017}.

\end{thebibliography}
